# Growth Phase Map of α- and β-Tantalum as a Function of Growth Temperature and Oxygen Partial Pressure

Marius O. Eji[1], Tyler J. Kirkpatrick[1], Ty J. Cawein[1], Joseph P. Corbett[1]

1. Miami University, Department of Physics, Oxford, Ohio 45056, USA

**Abstract**

Tantalum thin films are used in low-loss superconducting quantum circuits, yet the growth conditions governing selection of α-Ta(110), α-Ta(111), and metastable β-Ta on c-plane sapphire remain uncertain. We deposited 100 nm Ta films by dc magnetron sputtering on $Al_2O_3(0001)$ at 300–900 °C while varying the $O_2$ partial pressure from a $5.5 \times 10^{-11}$ Torr ultrahigh-vacuum baseline to $2.23 \times 10^{-4}$ Torr. Absorption-corrected X-ray diffraction texture coefficients were used to construct a ternary phase-orientation growth map in terms of temperature and the oxygen-to-Ta impingement flux ratio ($2\Phi_{O_2}/\Phi_{Ta} \leq 4.8 \times 10^{-6}$ to 19). Without added oxygen, α-Ta(110) dominated at every temperature. At an $O_2$ pressure of only $2.5 \times 10^{-9}$ Torr ($2\Phi_{O_2}/\Phi_{Ta} = 2.2 \times 10^{-4}$), α-Ta(111) accounted for 98–99% of the normalized texture weight from 300 to 700 °C, decreasing to 65% at 900 °C. Substantial β-Ta formed only at the highest oxygen dose, dominating at 300 °C and disappearing at 900 °C. The apparent β(001)/α(110) selection energy of $1.07 \pm 0.16$ eV is comparable to the activation energy for oxygen diffusion in Ta, consistent with oxygen transport influencing phase selection. Oxygen also produced orientation-dependent changes in crystallite size and microstrain. These findings establish oxygen-controlled processing windows for phase- and orientation-selective growth of Ta films for superconducting devices.

# I. INTRODUCTION

Tantalum (Ta) often thought of in the context of a refractory metal with a high melting point of 3017 °C [1] used in protective coatings [2] and has emerged as a leading material for low-loss superconducting circuits [3–15]. Replacing Nb with Ta in two-dimensional transmon qubits increased both lifetimes and coherence times beyond 0.3 ms [9]. Subsequent Ta transmons reached lifetimes of 0.5 ms [8] and, on high-resistivity Si, 1.68 ms, while Ta-based on-chip quantum memories achieved energy-relaxation times of 1.0–1.4 ms [16]. In Ta circuits on sapphire, dielectric two-level systems (TLSs) dominate loss, with comparable contributions from surface and bulk dielectrics [13,17]. Accordingly, recent work has examined native-oxide chemistry [14], the metal–substrate interface, where excess microwave loss has been associated with an epitaxial Ta/sapphire interface [4], and film microstructure, whose influence remains less clear. For example, TLS loss was higher in Nb-nucleated than in non-nucleated α-Ta(110) films on c-plane sapphire, approximately following differences in defect density inferred from the (110) reflection width and domain size [5]. In contrast, devices on c-plane sapphire with reported grain areas of 924 and 1700 nm² showed no statistically significant difference in performance [7]. Similarly, α-Ta grown on Nb-seeded Si at room temperature produced resonator quality factors comparable to those of high-temperature-grown films despite a substantially higher density of oxygen-rich grain boundaries [3]. These contrasting findings motivate closer examination of the growth conditions that establish the Ta interface and microstructure.

Ta thin films commonly exhibit two crystalline phases (**Figure 1**). The α phase is body-centered cubic (bcc) ($Im\bar{3}m$; a ≈ 0.331 nm) [10]. Films grown on Nb underlayers thicker than 0.3 nm superconduct near 4.3 K and exhibit a low-temperature resistivity of approximately 6 μΩ cm [18]. For α-Ta(111) on c-plane sapphire, the critical temperature decreases from 4.2 K at a film thickness of 269 nm to 2.9 K at 7.5 nm [10]. The metastable β phase, first reported in Ta thin films by Read and Altman [19] and frequently observed alongside α-Ta in sputtered films [20,21], adopts a tetragonal σ-type Frank–Kasper structure ($P\bar{4}2_1m$; a = 1.0211 nm, c = 0.5306 nm; 30 atoms per unit cell) [22–24]. β-Ta exhibits a higher resistivity of 150–170 μΩ cm [18,25] and does not superconduct above 1 K [18]. Nevertheless, it has applications beyond conventional superconducting circuitry: β-Ta is approximately twice as hard as α-Ta and has been proposed for wear-resistant biomedical coatings [1]; its large spin Hall effect enables switching of an adjacent ferromagnet [26]; and it has been used as the inductive element in hybrid kinetic-inductance detectors [27]. More recently, β-Ta films on sapphire have yielded transmon qubits with a mean time-averaged quality factor of $5.6 \times 10^6$ [28]. Understanding the conditions that determine Ta phase and α-Ta crystallographic orientation is therefore important for tailoring films to superconducting devices.

C-plane sapphire ($Al_2O_3(0001)$) is widely used as a substrate for Ta films in superconducting qubits [11,14,29], yet understanding what drives phase selection and crystallographic orientation remain unresolved. Direct deposition of α-Ta has generally required substrate temperatures above 400 °C [9,15], consistent with the temperature at which β-Ta is no longer observed in films sputtered on Si [21], or the use of seed and buffer layers such as Nb [3,5,18] and $TiN_x$ [30]. However, α-Ta has also been grown on $Al_2O_3(0001)$ by molecular beam epitaxy below 20 K, with the β-Ta fraction increasing at higher growth temperatures [29]. Similarly, α-Ta can be deposited on Si at 200 °C when Kr replaces Ar as the sputtering gas [15]. These contrasting results indicate that substrate temperature alone does not determine phase selection.

Beyond phase selection, the preferred orientation of α-Ta on c-plane sapphire remains incompletely understood. Sputtered films used in superconducting qubits commonly exhibit a dominant α-Ta(110) orientation, with occasional α-Ta(111) and β-Ta grains [11,29]. α-Ta(110)-oriented films have been reported at growth temperatures of 500 °C [5] and as high as 1150 °C[1]. In the latter case, the films exhibited a single out-of-plane orientation and three in-plane rotational variants on $Al_2O_3(0001)$. Although both α-Ta(110) and α-Ta(111) have been observed on this substrate, the growth conditions favoring one orientation over the other have not been established. Because microwave loss is sensitive to the Ta/sapphire interface [4], identifying the growth parameters that control orientation at this interface may provide an additional route to improving device performance.

The formation of metastable β-Ta thin films has been attributed to three interrelated factors: impurities, substrate-dependent nucleation, and deposition conditions. Surface hydroxyls and other impurities influence whether α- or β-Ta nucleates [31,32], while oxygen incorporated into β-Ta inhibits transformation to α-Ta and increases the temperature required for complete conversion [33]. Substrate selection also affects nucleation. At room temperature, Ta grows as β-Ta on Si but as α-Ta on Nb underlayers thicker than 0.3 nm [18]. On Si, β-Ta nucleation has been attributed to the formation of an amorphous Ta(Si) interfacial layer that minimizes interfacial energy [34]. In another study, α-Ta developed only after a 7–10 nm β-Ta interlayer formed, with substrate Debye temperature identified as the determining factor [35]. Deposition conditions provide a third influence: incident ion energy and the ion-to-Ta flux ratio affect phase selection [36], as do sputtering-gas mass [15] and energetic processes including ion-assisted, high-power impulse, and deep-oscillation magnetron sputtering [37–41]. Sputtering pressure also affects β-Ta formation, texture, and residual stress [25].

These factors are coupled, and the role of oxygen remains ambiguous. Although oxygen influences β-Ta stability [25,33], phase-pure β-Ta films have been produced under ultrahigh-vacuum conditions with low oxygen partial pressures [25,42]. Furthermore, impurity content was not found to determine phase selection for films sputtered on Si [34]. Structure zone models provide a framework for examining these competing influences on film microstructure. Conventional models describe microstructural evolution in terms of homologous temperature, $T_h = T/T_m$, while extended formulations account for energy delivered to the growth surface [43] and impurities in the depositing vapor, which can modify nucleation, coalescence, and grain growth [44,45].

The effect of oxygen on Ta phase and orientation selection remains unresolved because prior studies have examined different stages or environments of film growth. Controlled oxygen dosing has been used to investigate the β→α transformation in films initially deposited as β-Ta [33], and oxygen-rich grain boundaries have been observed in α-Ta [3]. Neither observation directly establishes how oxygen affects phase selection or the microstructure of competing orientations during deposition. In addition, the conclusion that impurity content is not decisive was reached for films grown on Si, where β-Ta nucleates on an amorphous interfacial layer [34]. Growth on c-plane sapphire differs because α-Ta develops epitaxial relationships with a crystalline substrate [5,6,12]. Finally, no growth variable has, to our knowledge, been shown to reproducibly switch α-Ta between the (110) and (111) orientations on c-plane sapphire. Variations in residual oxygen partial pressure among deposition systems could contribute to differences in reported orientations, but this possibility has not been systematically tested. A controlled map of Ta phase, orientation, and microstructure over a broad range of oxygen partial pressures, independently of substrate temperature, is therefore needed.

Here, we isolate oxygen exposure as a growth variable by depositing 100 nm Ta films on c-plane sapphire using direct current (DC) magnetron sputtering. Substrate temperatures range from 300 to 900 °C (homologous temperature $T_h = 0.17–0.36$, using $T_m = 3290$ K [46]), while $O_2$ partial pressure is varied from the ultrahigh-vacuum base pressure of $5.5 \times 10^{-11}$ Torr to $2.23 \times 10^{-4}$ Torr. We quantify oxygen exposure by the incident oxygen-to-Ta atomic flux ratio, $2\Phi_{O_2}/\Phi_{Ta}$, which spans a background upper bound of $4.8 \times 10^{-6}$ to a maximum of 19, corresponding to 19 incident O atoms per deposited Ta atom. Phase and preferred orientation are quantified using texture coefficients referenced to a calculated combined α-Ta and β-Ta powder pattern and corrected for thin-film absorption, placing reflections from both phases on a common intensity scale. Out-of-plane crystallite size and microstrain are determined from X-ray diffraction line-profile analysis, providing structural measures relevant to the reported association between α-Ta(110) domain structure and TLS loss [5]. The results are presented as a kinetic process map, rather than an equilibrium phase diagram, to identify growth conditions for controlling phase and orientation.

Without intentional oxygen dosing or a seed layer, α-Ta grows with a dominant (110) orientation throughout the 300–900 °C range. Introducing trace oxygen ($2\Phi_{O_2}/\Phi_{Ta} \approx 2 \times 10^{-4}$; $O_2$ partial pressure ≈ $2.5 \times 10^{-9}$ Torr) instead favors α-Ta(111), which becomes the dominant orientation from 300 to 700 °C. This preference weakens at 900 °C, consistent with increased grain-boundary mobility. β-Ta is absent at 900 °C at every oxygen dose and becomes dominant only at the highest dose. The temperature dependence of its competition with α-Ta(110) yields an apparent selection energy of 1.07 ± 0.16 eV, comparable to the reported activation energy for oxygen diffusion in Ta [47]. At 300 °C and an intermediate oxygen-to-Ta flux ratio of $2\Phi_{O_2}/\Phi_{Ta} = 0.09$, α-Ta(111) exhibits smaller out-of-plane crystallites and increased microstrain, whereas the α-Ta(110) crystallite size increases. This contrasting behavior suggests preferential oxygen retention within or near α-Ta(111) grains. The sensitivity of orientation selection to $O_2$ partial pressures below $10^{-8}$ Torr identifies trace oxygen as a possible source of variability among prior studies and establishes a process window for phase- and orientation-selective Ta growth relevant to superconducting devices.

# II. EXPERIMENTAL

## *A. Thin Film Growth*

Tantalum thin films were deposited by DC magnetron sputtering onto single-crystal c-plane sapphire substrates ($Al_2O_3$(0001); AdValue Technology Inc.) in a custom ultrahigh-vacuum (UHV) sputter-epitaxy system with a base pressure of $1.0 \times 10^{-10}$ Torr. Before deposition, the substrates were annealed at 1200 °C for 24 h under UHV while an in situ ultraviolet lamp (RBD Inc.) illuminated the heater stage, substrate, and chamber to promote water desorption. The chamber pressure remained at or below $3.0 \times 10^{-10}$ Torr during annealing. Substrate temperatures of 300, 500, 700, and 900 °C correspond to homologous temperatures $T_h = T/T_m$ of 0.17, 0.24, 0.30, and 0.36, respectively, using $T_m = 3290$ K for Ta [46]. The substrates were heated in direct contact with a commercial high-temperature heater stage (Bricada Inc.), with the stage temperature calibrated using a thermocouple in direct contact with the heater.

For each deposition, the oxygen background was established before introducing Ar. Five intentional $O_2$ partial pressures ($2.5 \times 10^{-9}$, $4 \times 10^{-8}$, $2.2 \times 10^{-7}$, $1 \times 10^{-6}$, and $2.23 \times 10^{-4}$ Torr) were established using a UHV-compatible variable leak valve, together with a no-added-$O_2$ condition corresponding to a residual $O_2$ partial pressure of $5.5 \times 10^{-11}$ Torr. The partial pressures were monitored using an Extorr 100 residual gas analyzer (RGA) before Ar admission. Ultrapure Ar was then introduced through a separate UHV-compatible variable leak valve to a working pressure of $3.2 \times 10^{-2}$ Torr. A 1.33 in. diameter Ta target (99.95% purity; Plasma Materials) was sputtered at a constant DC power of 100 W for 67 s. Film thickness was approximately 100 nm, corresponding to a deposition rate of 1.49 nm $s^{-1}$ determined from AFM step-height measurements on lift-off calibration specimens. To assess contaminants introduced by the Ar supply, Ar was admitted gradually to $1 \times 10^{-5}$ Torr, below the stated $1 \times 10^{-4}$ Torr RGA operating-pressure limit for Ar. No measurable increase in the $O_2$ ($m/z = 32$) or $H_2O$ ($m/z = 18$) signals was observed during this check. The Ar working pressure, sputtering power, and deposition time were held constant; only substrate temperature and oxygen exposure were varied.

The measured $O_2$ partial pressure was used to quantify the incident oxygen supply rather than the oxygen concentration incorporated into the film. The $O_2$ molecular impingement flux was calculated from kinetic gas theory as $\Phi_{O_2} = P(O_2)/\sqrt{[2\pi m(O_2)k_B T]}$, which at a gas temperature of 300 K gives $\Phi O_2 \approx (3.58 \times 10^{20}$ cm$^{-2}$ s$^{-1}$ Torr$^{-1}$)P($O_2$) [48]. The Ta deposition flux was calculated as $\Phi_{Ta} = \rho N_a r/M = 8.27 \times 10^{15}$ atoms cm$^{-2}$ s$^{-1}$, using the Ta density $\rho = 16.65$ g cm$^{-3}$, molar mass M = 180.95 g mol$^{-1}$, deposition rate r = 1.49 nm s$^{-1}$, and Avogadro constant $N_a$. The ratio $2\Phi_{O_2}/\Phi_{Ta}$ represents the number of incident oxygen atoms per deposited Ta atom. Across the six oxygen conditions, this ratio ranged from $\leq 4.8 \times 10^{-6}$ for the residual background to 19.3 for the highest intentional $O_2$ dose (Table I). The incident-flux ratio is an upper bound on incorporation from molecular oxygen if the arriving oxygen has a sticking probability no greater than unity; it is not a measurement of film oxygen stoichiometry. Even at the highest dose, the deposited films remained crystalline Ta rather than forming tantalum-oxide films.

**TABLE I.** Oxygen partial pressure, calculated molecular impingement flux, and oxygen-to-Ta atomic flux ratio for six growth conditions ($\Phi Ta = 8.27 \times 10^{15}$ atoms cm$^{-2}$ s$^{-1}$).

| $P_{O_2}$ (Torr) | $\Phi_{O_2}$ (cm$^{-2}$ s$^{-1}$) | $2\Phi_{O_2}/\Phi_{Ta}$ | $\log_{10}(2\Phi_{O_2}/\Phi_{Ta})$ |
|---|---|---|---|
| $5.5\times10^{-11}$ (base; no added $O_2$) | $1.97\times10^{10}$ | $\leq 4.8\times10^{-6}$ | $\leq -5.32$ |
| $2.5\times10^{-9}$ | $8.95\times10^{11}$ | $2.2\times10^{-4}$ | −3.66 |
| $4\times10^{-8}$ | $1.43\times10^{13}$ | $3.5\times10^{-3}$ | −2.46 |
| $2.2\times10^{-7}$ | $7.88\times10^{13}$ | 0.019 | −1.72 |
| $1\times10^{-6}$ | $3.58\times10^{14}$ | 0.087 | −1.06 |
| $2.23\times10^{-4}$ | $7.98\times10^{16}$ | 19.3 | +1.29 |

## B. *X-Ray Diffraction and Analysis*

### 1. *X-Ray Diffraction Measurement*

X-ray diffraction (XRD) patterns were acquired on a Bruker D8 Advance diffractometer with Cu Kα radiation ($K\alpha_1$, $\lambda = 1.54060$ Å; $K\alpha_2$, $\lambda = 1.54443$ Å) in parallel-beam geometry. Symmetric, coupled θ–2θ scans were collected over 2θ = 4°–120° in 0.02° steps with a counting time of 1 s per step. Diffraction peaks were analyzed using Fityk [49], and theoretical powder intensities were calculated using PowderCell [50] for the texture analysis.

In this symmetric geometry, the scattering vector is normal to the substrate surface, so the measured reflections correspond to lattice planes parallel to the substrate surface and probe the preferred out-of-plane orientation of each phase [51]. Instrumental broadening was characterized using a sapphire (corundum) powder standard under the same measurement configuration across the angular range. The estimated precision of the interplanar spacings was ±0.01 Å.

Each film reflection was fitted with a Voigt profile accounting for the Cu $K\alpha_1/K\alpha_2$ doublet [51]. Peak positions, integrated intensities, and Gaussian and Lorentzian integral breadths were extracted for the texture and diffraction line-profile analyses.

### 2. *X-Ray Line Analysis: Texture Coefficients*

Measured integrated diffraction intensities cannot be compared directly across reflections or phases because they depend on structure factor, multiplicity, Lorentz–polarization effects, and the angle-dependent diffracting volume of a thin film [51]. Texture coefficients account for these differences by normalizing each measured integrated intensity to that of the corresponding reflection from an isotropic reference [52–57]. Reference intensities $I_0$ were calculated using PowderCell [50] for a randomly oriented 50:50 mixture by mass of α-Ta ($Im\bar{3}m$; a = 3.30986 Å) and β-Ta ($P\bar{4}2_1m$; a = b = 10.2110 Å, c = 5.30640 Å) [23], placing both phases on a common reference intensity scale. Because the film thickness (approximately 100 nm) is substantially smaller than the X-ray attenuation length in Ta at Cu Kα wavelengths, the diffracting volume depends on Bragg angle [51]. Each reference intensity was therefore adjusted using the thin-film absorption factor $G_{hkl} = 1 - \exp(-2\mu t/\sin \theta_{hkl})$ for a flat film of thickness (t) in symmetric geometry. The linear attenuation coefficient (μ) was evaluated using the NIST mass attenuation coefficient $\mu/\rho \approx 161$ cm$^2$ g$^{-1}$ and density (ρ) = 16.65 g cm$^{-3}$, giving an attenuation length $1/\mu \approx 3.7$ μm [58]. The texture coefficient ($C_i$) for reflection i, specified by its phase and Miller indices, is defined as

$$C_i = \frac{I_i/I_{0,i}}{\frac{1}{N}\sum_j I_j\,/I_{0,j}} \qquad (1)$$

where $I_i$ is the integrated intensity obtained from the Voigt-doublet fit. The sum includes a fixed set of N = 153 candidate reflections from α-Ta and β-Ta within the measured 2θ range. At most five reflections were observed in an individual film: α-Ta(110), α-Ta(220), α-Ta(222), β-Ta(002), and β-Ta(004). Reflections below the detection limit were assigned zero integrated intensity, and therefore a texture coefficient of zero. Because all coefficients for a given film share the same normalization sum, changing N rescales them by a common factor; coefficient ratios and the ternary fractions derived from them are independent of N.

By definition, $\Sigma C_i = N$, and each reflection of the randomly oriented 50:50 α-Ta/β-Ta reference has $C_i = 1$. Values greater than unity indicate enhancement relative to the isotropic reference, whereas values between zero and unity indicate reduced relative intensity [55]. When only one out-of-plane pole family is observed, its reflections account for the entire normalization sum. For example, $C_{\alpha 222} = N$ when α-Ta{111} is represented only by the (222) reflection, whereas $C_{\alpha 110} + C_{\alpha 220} = N$ when only the α-Ta{110} family is observed. Such a result indicates a single detectable out-of-plane orientation, not a single-crystalline film because symmetric θ–2θ scans do not establish in-plane alignment or mosaic spread. The coefficients also depend on the contributions of different phases and their orientation distributions. Differences in mosaic spread can bias the relative weight of orientation components compared with their volume fractions; additional rocking-curve measurements would be required to evaluate these effects.

The α-Ta(110)/(220) and β-Ta(002)/(004) pairs probe the same out-of-plane pole families. Their mean texture coefficients were used to represent the α-Ta{110} and β-Ta{001} families, respectively: $P_{\alpha 110} = (C_{\alpha 110} + C_{\alpha 220})/2$ and $P_{\beta 001} = (C_{\beta 002} + C_{\beta 004})/2$. The expected ratio of coefficients for each pair is unity, providing an internal check on peak integration, reference intensities, and absorption corrections. Across the measured films, $C_{\alpha 220}/C_{\alpha 110} = 1.03 \pm 0.25$. We therefore assigned a relative uncertainty of ±25% to each texture coefficient. Although the fitted integrated-area uncertainties were below 10% for the strongest reflections, they were larger for weak peaks. The ±25% estimate instead reflects the observed scatter between equivalent pole-family measurements and includes contributions from background subtraction, absorption correction, and reference-intensity calculations.

### 3. *X-Ray Line Analysis: Crystallite Size and Microstrain*

Diffraction peak broadening was analyzed to estimate the out-of-plane coherent domain size and lattice microstrain. The measured breadth contains contributions from finite crystallite size, nonuniform strain, and instrumental resolution. Reflections were represented using Voigt profiles, which combine Gaussian and Lorentzian broadening. The fitted Gaussian and Lorentzian full widths at half maximum (FWHM) were converted to integral breadths using $\beta_G = 1.0645$ $FWHM_G$ and $\beta_L = (\pi/2)FWHM_L$ [59]. The instrumental profile was measured with a corundum powder standard in the same parallel-beam configuration, and the angular dependence of its Gaussian and Lorentzian components was fitted separately using second-order polynomials [60]. Instrumental broadening was subtracted linearly from the Lorentzian breadth and in quadrature from the Gaussian breadth [61]. In the single-line Voigt approximation, the corrected Lorentzian component was attributed to coherent domain size (D), and the Gaussian component to apparent microstrain ($\varepsilon$) [61]. The corresponding Scherrer and strain relations are

$$D = \frac{K\lambda}{\beta_L \cos\theta}, \qquad \varepsilon = \frac{\beta_G}{4\tan\theta} \qquad (2)$$

where $\lambda$ is the Cu $K\alpha_1$ wavelength, $\theta$ is the Bragg angle, and both breadths are expressed in radians of $2\theta$ [8,14,17–19]. With the shape factor $K = 0.95$, D represents the volume-weighted mean extent of coherently diffracting domains normal to the reflecting planes [62]. Because the reflecting planes are parallel to the substrate in symmetric geometry, D measures the out-of-plane coherent domain size. The dimensionless parameter $\varepsilon$ is an apparent upper-limit estimate of microstrain and is reported in units of $10^{-3}$ [9,18,19]. Size and strain were evaluated separately for α-Ta(110) using the (110) reflection, α-Ta(111) using the (222) reflection, and β-Ta(001) using the (002) reflection. Because separating size and strain from a single diffraction line depends on the assumed peak-shape model [61,65], the analysis emphasizes relative trends across growth conditions rather than absolute values. Lower-angle reflections, such as α-Ta(110), are comparatively more sensitive to size broadening, whereas higher-angle reflections, such as α-Ta(222), are more sensitive to strain broadening. Consequently, the uncertainties of the two α-Ta orientation components need not be identical.

## C. *Atomic Force Microscopy*

Film surface morphology was characterized by atomic force microscopy (AFM) using a high-resolution multimodal instrument (AFM Workshop). Pt-coated cantilevers (Budget Sensors; nominal resonance frequency 75 kHz and spring constant 3 N $m^{-1}$) were used for all measurements. Samples were measured in a nitrogen-filled glovebox (MSE Inc.) equipped with a load lock for transfer. AFM images were processed using Gwyddion [66]. Individual grains were identified using the software's segmentation masks and approximated by ellipses to determine their major and minor axes and the in-plane orientation of the major axis. Finite tip size broadens the apparent lateral extent of surface protrusions and reduces their apparent separation [67]; therefore, the reported lateral grain dimensions should be regarded as apparent values, particularly for features comparable in size to the tip radius.

# III. RESULTS

## A. *XRD Symmetric Coupled Scans*

**Error! Reference source not found.** depicts symmetric coupled θ–2θ XRD scans for the 24 films. Across the investigated growth conditions, only five Ta reflections are detected: α-Ta (110), (220), and (222) at 2θ ≈ 38.4°, 82.2°, and 108.2°, respectively, and β-Ta (002) and (004) at 2θ ≈ 33.7° and 71.0°, respectively. The (111) reflection is forbidden for body-centered-cubic α-Ta because h + k + l is odd; accordingly, the (111) out-of-plane orientation is identified by its (222) reflection. Without intentional oxygen dosing ($2\Phi_{O_2}/\Phi_{Ta} \leq 4.8 \times 10^{-6}$), only α-Ta (110) and (220) reflections are observed at all four growth temperatures. At the two lowest added oxygen flux ratios ($2.2 \times 10^{-4}$ and $3.5 \times 10^{-3}$), the α-Ta (222) reflection appears alongside the (110) family throughout the temperature series. Weak β-Ta (002) reflections are additionally detected at 300 °C under these two conditions.

At intermediate oxygen flux ratios of 0.019 and 0.087, α-Ta (110) and (222) reflections generally coexist across the temperature series; the exception is the absence of the (222) reflection at 500 °C. At the highest oxygen flux ratio ($2\Phi_{O_2}/\Phi_{Ta} = 19$), β-Ta (002) and (004) reflections are observed at 300, 500, and 700 °C, whereas the α-Ta (222) reflection is absent. Only β-Ta reflections are detected at 300 °C under this highest-dose condition. Because β-Ta can also nucleate in the absence of oxygen when favored by the substrate interface [68], these observations identify oxygen as a variable controlling β-Ta formation under the present deposition conditions, not as a necessary cause of β-Ta nucleation. No β-Ta reflections are detected at 900 °C at any oxygen dose. These scans establish the phases and out-of-plane orientations present; their relative texture weights are evaluated below rather than inferred directly from raw peak intensities.

Interplanar spacings were determined from the fitted Cu $K\alpha_1$ peak positions. Across the growth series, α-Ta exhibits d-spacings of 2.34–2.35 Å for (110), 1.170–1.172 Å for (220), and 0.95–0.96 Å for (222), whereas β-Ta exhibits 2.65–2.66 Å for (002) and 1.3266–1.3270 Å for (004). The corresponding maps of α-Ta (110), α-Ta (222), and β-Ta (002) d-spacings are shown in Error! Reference source not found.. These values agree with the spacings expected for α-Ta with $a \approx 0.331$ nm [10] and β-Ta with $c = 0.53064$ nm [23]. Within the measurement precision of ±0.01 Å (approximately 0.4% for the α-Ta (110) reflection and 1% for (222)), no systematic dependence of d-spacing on temperature, oxygen flux ratio, or phase coexistence is resolved. Consequently, differences in out-of-plane lattice strain between the α-Ta (110) and (111) components cannot be established, and smaller uniform changes in lattice spacing remain below the measurement resolution.

## B. *Texture Coefficient Analysis*

**Error! Reference source not found.** presents the texture coefficients calculated from the integrated XRD peak areas using Eq. (1). The five observed reflections are distinguished by color: α-Ta (110) and (220) in two shades of blue, α-Ta (222) in yellow, and β-Ta (002) and (004) in pink and red, respectively. For each film, the coefficients sum to N = 153 over the fixed reference reflection set. Because at most five reflections are detected, strongly preferred components carry coefficients well above the isotropic-reference value of C = 1, typically exceeding 30. None of the films approaches a random out-of-plane orientation distribution.

**Error! Reference source not found.** maps the texture coefficients of the three phase–orientation components across growth temperature and oxygen flux ratio. The color scheme matches that of the bar plots. Black tiles indicate weak detected reflections with C ≲ 1, which do not constitute a preferred orientation relative to the isotropic reference. Gray tiles indicate reflections that were not detected. This presentation separates strongly textured components from weakly represented or absent components without assigning the weak signals to a particular nucleation mechanism.

The maps reveal competition among the observed phase–orientation components. Conditions with weak α-Ta (110) texture, shown as dark tiles in **Error! Reference source not found.**(a), generally exhibit strong α-Ta (111) texture, indicated by the bright tiles in **Error! Reference source not found.**(b). Conversely, conditions with weak α-Ta (111) texture, as shown in **Error! Reference source not found.**(b), do not invariably exhibit the strongest α-Ta (110) texture because β-Ta may also contribute. At the highest oxygen dose, β-Ta texture is strongest where α-Ta is absent and decreases with increasing deposition temperature as α-Ta (110) becomes more prominent, as shown in **Error! Reference source not found.**(c).

## C. *Growth Phase Maps*

The texture coefficient maps provide the basis for a growth phase–orientation map expressed using parameters motivated by structure zone models. The temperature coordinate is the homologous temperature, $T_h = T/T_m$, which represents the thermal contribution in the extended structure zone framework of Anders. DC sputtering power, Ar working pressure, and nominal film thickness were held constant, fixing the corresponding deposition conditions across the experimental series [43]. The investigated range ($T_h$ = 0.17–0.36) extends from growth conditions with limited adatom mobility and persistent nucleation-selected textures toward conditions in which grain-boundary migration becomes increasingly important [45,69]. For bcc metals, the close-packed (110) planes are generally expected to be favored because of their relatively low surface energy [70,71]. The oxygen-induced preference for α-Ta (111) observed here therefore departs from this simple surface-energy expectation. The second coordinate is the oxygen-to-Ta impingement flux ratio, $2\Phi_{O_2}/\Phi_{Ta}$, motivated by impurity-dependent extensions of structure zone models in which impurities can modify nucleation, coalescence, and grain growth [44]. The use of homologous temperature and a normalized oxygen flux ratio facilitates comparison with growth processes performed in other deposition systems.

For each film, the family texture coefficients $P_{\alpha110}$, $P_{\alpha111}$, and $P_{\beta001}$ were normalized to three weights (a, b, c) summing to unity. The resulting ternary diagram has vertices corresponding to α-Ta (110), α-Ta (111), and β-Ta (001). All measured coefficients were retained without imposing a texture threshold. Each film is represented by a tile colored according to its ternary coordinates; the vertex colors are mixed in OKLab color space to distinguish intermediate compositions, with the ternary diagram providing the color key. Importantly, these weights represent phase–orientation contributions to the measured out-of-plane pole densities, not independently measured phase volume fractions. Gray at the ternary center denotes equal weights of the three components.

Error! Reference source not found. summarizes the resulting phase–orientation map. Without intentional oxygen dosing, α-Ta (110) is the dominant component at every growth temperature. A small added oxygen flux reverses this preference: at $2\Phi_{O_2}/\Phi_{Ta} = 2.2 \times 10^{-4}$ (approximately two impinging O atoms per $10^4$ deposited Ta atoms), α-Ta (111) accounts for 98–99% of the ternary weight from 300 to 700 °C and 65% at 900 °C. Excluding trace contributions, β-Ta is detected only at the highest oxygen flux ratio of 19. Under that condition, β-Ta is the sole detected phase at 300 °C, dominates the ternary weight at 500 °C ($c = 0.66$), and becomes a minor component at 700 °C ($c = 0.07$). No β-Ta is detected at 900 °C. The only other β-Ta detections occur at 300 °C for the two lowest added oxygen flux ratios, where $c \leq 0.9$. Orientation selection at intermediate doses is nonmonotonic, with α-Ta (110) again becoming dominant at 300 °C and 0.019; 500 °C and 0.019 or 0.087; and 700 °C and $3.5 \times 10^{-3}$. These local reversals are retained in the map, while the strongest systematic features are the response to trace oxygen and the temperature-dependent β-Ta region at high oxygen dose. At 900 °C, four of the six films are α-Ta (110)-dominant; the two lowest added oxygen doses instead favor α-Ta (111), with 65–78% of the ternary weight.

## D. *Crystallite Size and Microstrain*

Error! Reference source not found. presents the out-of-plane crystallite size $D$ and apparent microstrain $\varepsilon$ of α-Ta (110), α-Ta (111), and β-Ta (001) on the growth coordinates used in **Error! Reference source not found.**. We assign uncertainties of ±20 nm to D and $\pm 0.75 \times 10^{-3}$ to $\varepsilon$. Single-line integral-breadth analysis has reported relative uncertainties of approximately one-third in crystallite size even when individual peaks are well fitted[20]. The assigned 20 nm uncertainty is also comparable to the standard deviation of D across the map; thus, differences such as 35 nm versus 45 nm are not resolved. Values approaching the nominal 100 nm film thickness are consistent with coherent diffracting domains extending through much of the film thickness.

Error! Reference source not found. (a,c,e) shows the microstrain distributions. The smallest values for the two α-Ta orientations occur at relatively low growth temperature and oxygen dose: $\varepsilon \approx 0.4 \times 10^{-3}$ for α-Ta (110) at 300 °C without added oxygen and for α-Ta (111) at 500 °C without added oxygen. At 300 °C, the orientations reach their maximum measured microstrain at different doses. For α-Ta (110), $\varepsilon = 4.0 \times 10^{-3}$ at $2\Phi_{O_2}/\Phi_{Ta} = 3.5 \times 10^{-3}$; for α-Ta (111), $\varepsilon = 93 \times 10^{-3}$ at $2\Phi_{O_2}/\Phi_{Ta} = 0.087$. Outside these elevated values, the microstrain is comparatively insensitive to the growth variables, with the largest variations concentrated at lower temperature and higher oxygen exposure. β-Ta (001) reaches $\varepsilon = 3.9 \times 10^{-3}$ when present only as a trace component within an α-Ta-dominated film at 300 °C ($2\Phi_{O_2}/\Phi_{Ta} = 2.2 \times 10^{-4}$ and $3.5 \times 10^{-3}$). At the highest oxygen dose, its microstrain is approximately constant within uncertainty ($\varepsilon \approx 2.1 \times 10^{-3}$) from 300 to 700 °C.

Crystallite size, Error! Reference source not found. (b,d,f) does not exhibit a simple correlation with texture coefficient. The α-Ta (110) component has a mean D of 47 ± 20 nm across the map, with no resolvable variation among the films grown at 900 °C. Its largest value, 116 ± 20 nm, occurs at 300 °C and $2\Phi_{O_2}/\Phi_{Ta} = 0.087$ and is comparable to the film thickness within uncertainty. All remaining α-Ta (110) values lie within approximately 1.3 standard deviations of the 47 nm mean. For α-Ta (111), the largest D values (70 ± 20 nm) occur at 500 °C and $2.2 \times 10^{-4}$ and at 700 °C and 0.087. These conditions define a trend toward larger coherent domains with increasing temperature and oxygen dose. The smallest reported value is 7 ± 20 nm at 900 °C and 0.019, for which the uncertainty precludes a precise size estimate.

The greatest contrast between the α-Ta orientations occurs at 300 °C and $2\Phi_{O_2}/\Phi_{Ta} = 0.087$. Under these conditions, α-Ta (110) has its largest measured crystallite size (D = 116 nm), whereas α-Ta (111) exhibits its largest apparent microstrain ($\varepsilon = 93 \times 10^{-3}$) and a crystallite size of 20 nm. Because the Gaussian and Lorentzian contributions are extracted from the same Voigt fit, covariance between the fitted breadths can produce correlated changes in the derived size and strain. The simultaneous decrease in D and increase in ε for α-Ta (111) opposes that expected covariance and therefore is unlikely to arise solely from this fitting effect. β-Ta exhibits a different relation between size and texture weight. At 300 °C, its trace contributions at $2\Phi_{O_2}/\Phi_{Ta} = 2.2 \times 10^{-4}$ and $3.5 \times 10^{-3}$ have D = 128 ± 20 nm and 105 ± 20 nm, respectively, comparable to the film thickness within uncertainty. At the highest oxygen dose, where β-Ta becomes prominent, D is 38, 39, and 81 nm at 300, 500, and 700 °C, respectively. No systematic dependence of D on texture coefficient is resolved for the α-Ta components; for β-Ta, larger coherent domains tend to accompany smaller texture coefficients.

### *E. Temperature Dependence of Phase and Orientation Selection*

Because all texture coefficients for a film share the same normalization denominator, ratios of family coefficients cancel this factor. Changing the α:β reference weighting introduces only a temperature-independent multiplier. The temperature dependence of such a ratio can therefore be evaluated using an Arrhenius plot at fixed oxygen dose (Error! Reference source not found.). At the highest oxygen flux ratio ($2\Phi_{O_2}/\Phi_{Ta} = 19$), the mixed-phase films at 500 and 700 °C yield $\ln(P_{\beta 001}/P_{\alpha 110}) = 0.66$ and $-2.63$, respectively, corresponding to β-Ta ternary weights of 0.66 and 0.07. These two points give an apparent selection energy ΔE = 1.07 ± 0.16 eV, with β-Ta favored at lower temperature. The single-phase endpoints constrain this trend: the β-only film at 300 °C sets a lower bound on the ratio, whereas the α-only film at 900 °C sets an upper bound. For plotting these limits, the undetected component is assigned the integrated area of the weakest detected reflection; this substitution does not enter the texture analysis or fit. Extrapolation predicts logarithmic ratios of 6.24 ± 1.13 at 300 °C and −4.79 ± 0.63 at 900 °C. The measured bounds ($\geq$ 4.44 and $\leq -4.43$, respectively) are consistent with the fitted trend, subject to the detection threshold discussed above. The assumed ±25% uncertainty in each coefficient gives ±0.35 in each logarithmic ratio. The extracted energy is therefore an apparent selection scale based on two mixed-phase data points bracketed by detection limits.

At the trace oxygen flux ratio $2\Phi_{O_2}/\Phi_{Ta} = 2.2 \times 10^{-4}$, the α-Ta (111)-to-(110) texture ratio shows no resolvable Arrhenius dependence between 300 and 700 °C. $P_{\alpha 111}/P_{\alpha 110}$ remains approximately 130–200 (ln ratio ≈ 5), corresponding to $|\Delta E| \lesssim 0.1$ eV and as such no activation energy is determined. Between $T_h = 0.30$ and 0.36, the ratio falls to approximately 1.9, although α-Ta (111) still contributes 65% of the ternary weight at 900 °C. Relative to films grown without added oxygen, this trace dose increases $P_{\alpha 111}/P_{\alpha 110}$ by approximately $10^3$, $5 \times 10^3$, $\geq 10^4$, and $\geq 2 \times$ $10^2$ at 300, 500, 700, and 900 °C, respectively. Across oxygen doses, the α-Ta (111) preference is generally stronger at 300 °C than at 900 °C.

### F. *Surface Morphology*

**Error! Reference source not found.**(a) compares the surface morphologies of two end members of the growth map: a β-Ta film and an α-Ta (110) film. Both exhibit rounded to elongated grains, but differ in height variation, grain dimensions, and the distribution of in-plane grain orientations. Relative to the mean plane, β-Ta surface heights lie within approximately ±0.25 nm, compared with ±0.75 nm for α-Ta (110). Grain boundaries were segmented in Gwyddion, and the resulting regions were fitted with ellipses to determine distributions of major and minor axes. The fitted mean major and minor axes are 30 ± 8 and 22 ± 6 nm for α-Ta (110), compared with 23 ± 14 and 14 ± 9 nm for β-Ta (002). Thus, α-Ta (110) exhibits larger, more uniform, and more elongated surface grains than β-Ta.

The grain-axis orientation distributions provide a further distinction between the two surfaces (**Error! Reference source not found.**). For α-Ta (110), the fitted major-axis angles span approximately −85° to +85° and exhibit three preferred directions separated by about 60°. These directions are also apparent in the topographs. Because ellipse axes are invariant under a 180° rotation, three in-plane variants related by 120° rotations can appear as three directions separated by 60° in the axial-angle distribution. This pattern is consistent with the symmetry of c-plane sapphire and with previously reported in-plane variants of α-Ta (110) on this substrate [12]. By contrast, β-Ta exhibits a broad major-axis distribution spanning approximately −70° to +70° without resolved peaks. The present AFM analysis therefore reveals a substrate-symmetry-consistent directional texture for α-Ta (110), but no comparable in-plane alignment for β-Ta (001).

Comparison with the out-of-plane crystallite sizes reported in Sec. III D is consistent with a columnar grain morphology. For the imaged films, $D = 48 \pm 20$ nm for α-Ta (110) and 39 ± 20 nm for β-Ta (002) at 300 °C, exceeding the corresponding mean minor-axis dimensions determined from AFM. These measurements suggest greater grain extent normal to the substrate than along the minor in-plane axis, consistent with columnar growth at $T_h \leq 0.36$ [45].

## IV. DISCUSSION

The growth phase–orientation map reveals three distinct responses to oxygen: a transition from α-Ta (110) to α-Ta (111) at trace oxygen flux that is has a large substrate temperature window up to $T_h = 0.30$ (Fig. 5); the formation of β-Ta (001) at the highest oxygen flux, with a pronounced temperature dependence; and orientation-dependent changes in crystallite size and microstrain (Fig. 6). These observations are considered separately below. Two limitations are important throughout. First, oxygen is a variable controlling phase and orientation selection under the present growth conditions, but is not a necessary condition for β-Ta formation, which may also be favored by substrate-dependent nucleation [68]. Second, film oxygen concentrations were not measured. The ratio $2\Phi_{O_2}/\Phi_{Ta}$ quantifies the incident oxygen supply, not the incorporated concentration which the impinging $O_2$ provides an upper-bound estimate of incorporation from that gas under the assumption of unitary sticking coefficient for $O_2$.

### A. *Orientation Selection of α-Ta by Trace Oxygen*

Without intentional oxygen dosing, α-Ta grows predominantly with its (110) planes parallel to the substrate at every temperature. This behavior is consistent with competitive growth favoring the close-packed, low-surface-energy {110} planes of a bcc metal (Sec. III C). A simple elastic-strain-energy argument does not account for the oxygen-induced (111) preference. Ta is elastically anisotropic, with $C_{11} = 262$ GPa, $C_{12} = 156$ GPa, and $C_{44} = 82.6$ GPa [72], corresponding to a Zener anisotropy ratio of $2C_{44}/(C_{11} - C_{12}) = 1.56$. The biaxial modulus increases from 232 GPa for (100) to 315 GPa for (111). For comparable imposed biaxial strains, elastic-energy minimization would favor (100), which is not detected in these films [73]. The (111) preference is therefore more consistent with interfacial orientation selection during nucleation than with this simple elastic-energy picture. Both α-Ta orientations have previously been reported on c-plane sapphire, including predominantly (110)-oriented sputtered films containing occasional (111) grains [11,29], (110)-oriented films grown at 500 °C [36], and films with a single (110) out-of-plane orientation grown at 1150 °C [12], as well as (111)-oriented films [6,10]. Their interfacial geometries differ. The α-Ta(111) plane forms a hexagonal surface net with an in-plane periodicity of $\sqrt{2}a \approx 0.468$ nm, within approximately 1.7% of the sapphire lattice parameter a = 0.4759 nm [74]. This net shares the threefold rotational symmetry of c-plane sapphire. In contrast, the α-Ta(110) surface net is rectangular and can form symmetry-related in-plane variants on the threefold substrate. Three such variants have been reported for (110)-oriented α-Ta on c-plane sapphire and are consistent with the preferred grain-axis directions observed by AFM (Fig. 8) [12].

The central observation is that trace oxygen strongly favors α-Ta (111) despite the small incident oxygen dose. At $2\Phi_{O_2}/\Phi_{Ta} = 2.2 \times 10^{-4}$, the 67 s deposition receives approximately $1.2 \times 10^{14}$ O atoms $cm^{-2}$ from $O_2$ impingement, equivalent to 0.09 of the areal atomic density of a Ta(110) layer ($1.29 \times 10^{15}$ atoms $cm^{-2}$). Even with unit sticking, this deposition-stage supply corresponds to only $2.2 \times 10^{-4}$ O atoms per deposited Ta atom, or approximately 220 atomic ppm. The sapphire is also exposed to $O_2$ for approximately 5 min at the growth temperature before deposition while residual-gas spectra are acquired. At pressure $O_2 = 2.5 \times 10^{-9}$ Torr, this exposure is 0.75 L, corresponding to approximately $5.4 \times 10^{14}$ impinging O atoms $cm^{-2}$. This is equivalent to 0.4 of the Ta(110) layer areal density and is approximately 4.5 times the oxygen impingement during deposition. The initial substrate surface therefore experiences a larger oxygen impingement dose than the growing film, although its adsorbed oxygen coverage at the growth temperature is unknown. The sensitivity of the final 100 nm film orientation to such a small dose is consistent with oxygen modifying the earliest stages of nucleation or growth-front evolution rather than acting solely through bulk incorporation. Two results support this interpretation. First, at $2\Phi_{O_2}/\Phi_{Ta} = 2.2 \times 10^{-4}$, the $P_{\alpha111}/P_{\alpha110}$ texture ratio is approximately temperature-independent between 300 and 700 °C ($|\Delta E| \lesssim 0.1$ eV; Sec. III E), consistent with orientation selection during nucleation persisting through subsequent growth under conditions of limited grain-boundary mobility [45,69]. Second, the (111) preference weakens between $T_h = 0.30$ and 0.36, where increased grain-boundary mobility may allow the lower-surface-energy (110) orientation to grow competitively. A reduced oxygen coverage at the growth front at 900 °C, arising from faster oxygen redistribution or desorption, could also contribute. These possibilities cannot be distinguished with the present measurements. Neither the oxygen termination of the sapphire surface nor the atomic structure and energy of the Ta/sapphire interface were measured, so the mechanism responsible for preferential (111) nucleation remains unresolved. The nonmonotonic reversions to (110) at intermediate oxygen fluxes may reflect competition between the two orientations during nucleation and growth, but the present data do not isolate the controlling process.

## B. *Oxygen and the Formation of β-Ta*

Substantial β-Ta formation is restricted to the highest oxygen flux ratio, $2\Phi O_2/\Phi Ta = 19$. At this dose, β-Ta is the only phase detected at 300 °C and is absent at 900 °C, yielding an apparent β-Ta(001)/α-Ta(110) selection energy of $\Delta E = 1.07 \pm 0.16$ eV (Sec. III E). This value is comparable to the reported activation enthalpy of 1.10 eV for interstitial oxygen diffusion in Ta, determined by anelastic spectroscopy [47]. Oxygen diffusion in Ta has also been examined using relaxation measurements over several decades [75,76]. The agreement motivates a kinetic interpretation in which β-Ta formation is promoted when oxygen arrives at the growth surface faster than it can be redistributed into the film. In a simplified diffusion-limited picture, the oxygen retained near the growth front may scale inversely with the oxygen diffusivity $D_o = D_0 \exp[-E_a/(k_B T)]$, so that $1/D_o \propto \exp[E_a/(k_B T)]$. A phase-selection ratio that tracks this surface concentration would then exhibit an apparent activation energy comparable to the diffusion barrier. Using $D_0 = 2.03 \times 10^{-5}$ cm² s⁻¹ and $E_a = 1.10$ eV [47], extrapolated beyond the temperature range of the original measurements, the characteristic lattice-diffusion length $\sqrt{(D_o t)}$ over the 67 s deposition is approximately 5 nm at 300 °C, 0.1 μm at 500 °C, 0.5 μm at 700 °C, and 1.6 μm at 900 °C. These estimates place the transition from β-Ta-dominant to α-Ta-dominant growth near temperatures where the calculated diffusion length becomes comparable to or exceeds the nominal 100 nm film thickness. Oxygen redistribution is strongly limited at 300 °C, whereas it can extend through much of the film at higher temperatures. This comparison is qualitative: the estimates describe lattice diffusion, while grain-boundary diffusion may be faster, and material deposited earlier experiences a longer diffusion time. The proposed oxygen-retention mechanism is also consistent with reports that incorporated oxygen increases the temperature required for the β-to-α transformation [33].

Limited oxygen redistribution cannot by itself account for the formation of β-Ta. Even at 300 °C, where the calculated diffusion length is shortest, oxygen flux ratios up to 0.087 produce no more than trace β-Ta ($c \leq 0.9$). The onset of β-Ta-dominant growth therefore also requires sufficient oxygen supply, with the relevant transition lying between the two highest investigated flux ratios, 0.087 and 19. The correspondence between the apparent selection energy and the oxygen diffusion barrier is consistent with, but does not establish, a diffusion-controlled mechanism. The measured texture ratio reflects both nucleation and subsequent growth, and other thermally activated processes with comparable energy scales cannot be excluded. Furthermore, β-Ta can be stabilized by substrate-interface-controlled nucleation without intentional oxygen dosing [68]. Phase selection is consequently expected to depend on the combined effects of substrate interface, temperature, incident particle energetics, working-gas conditions, and oxygen-bearing species [33,36,68].

## C. *Oxygen, Crystallite Size, and Microstrain*

The crystallite-size and microstrain maps provide additional constraints on the influence of oxygen. Four observations are particularly relevant. First, the strongest oxygen-dependent changes in α-Ta microstrain occur at 300 and 500 °C, whereas the microstrain is comparatively insensitive to oxygen dose at 700 and 900 °C. Oxygen can dissolve interstitially in α-Ta before oxide formation [33,77]. A spatially inhomogeneous interstitial distribution could produce local variations in lattice spacing and hence diffraction broadening without a resolvable shift in the mean d-spacing. At 300 and 500 °C, the estimated oxygen diffusion length during deposition is smaller than or comparable to the film thickness, allowing oxygen concentration gradients to persist. At 700 and 900 °C, the longer diffusion lengths permit greater redistribution, potentially including segregation to grain boundaries. This interpretation is consistent with oxygen-enriched grain boundaries observed in α-Ta films [3]. Thus, the temperature dependence of microstrain could reflect a competition between oxygen delivery and redistribution, analogous to the proposed explanation for the β-Ta boundary. Direct measurements of oxygen concentration and spatial distribution would be required to test this interpretation.

Second, films grown without intentional oxygen dosing exhibit their lowest microstrain at 300 °C and larger values at 700 and 900 °C. This trend is opposite to that expected if thermally activated defect recovery were the only relevant process. A temperature-dependent contribution unrelated to intentional oxygen dosing is therefore plausible. One possibility is differential thermal contraction of Ta and sapphire during cooling, which increases with the temperature interval between growth and measurement [46,74]. Uniform biaxial strain would primarily shift diffraction peak positions rather than broaden them; however, grain-to-grain differences in strain accommodation or in-plane orientation could generate a distribution of out-of-plane lattice spacings and contribute to the measured microstrain.

Third, the two α-Ta orientations respond differently to increasing oxygen flux at 300 °C. The microstrain of α-Ta (110) reaches a maximum at an intermediate dose but decreases at $2\Phi_{O_2}/\Phi_{Ta} = 0.087$, where its out-of-plane crystallite size approaches the film thickness. In contrast, α-Ta (111) exhibits its highest microstrain and smallest crystallite size at this same oxygen flux ratio. These opposing trends are consistent with preferential oxygen retention in or near α-Ta (111) grains, which could limit their coherent growth and introduce nonuniform strain, while α-Ta (110) retains larger coherent domains. Since impurities can either inhibit or promote microstructural evolution [44], the data suggest orientation-dependent responses to oxygen rather than a uniform reduction in crystallite size across the film. The location and chemical state of the oxygen, however, have not been determined.

Fourth, no systematic correlation between crystallite size D and texture coefficient C is resolved for either α-Ta orientation. Thus, the oxygen-induced (111) texture preference in the growth map (Fig. 5) cannot be explained simply by an increase in the out-of-plane coherent-domain size of that component. The largest α-Ta (111) crystallites occur along a trend of increasing temperature and oxygen flux, potentially reflecting increased oxygen redistribution at higher temperatures. This trend is consistent with, but does not independently verify, a competition between oxygen-induced growth inhibition and thermally activated transport.

Notably, the largest α-Ta (110) crystallite size is comparable to the nominal film thickness and occurs at 300 °C and $2\Phi_{O_2}/\Phi_{Ta} = 0.087$ rather than under the highest-temperature, lowest-oxygen growth conditions. Long out-of-plane coherent domains of α-Ta (110) therefore do not require the highest deposition temperatures investigated.

The β-Ta(001) component exhibits a different relationship between texture weight and crystallite size. When β-Ta is detected only as a minor component within an α-Ta-dominant film at 300 °C, its coherent-domain size is comparable to the film thickness and its microstrain is relatively high. This behavior is consistent with isolated β-Ta domains constrained by a surrounding α-Ta matrix, although their spatial arrangement is not established by the symmetric XRD measurements. At the highest oxygen dose, where β-Ta is dominant at lower temperatures, its crystallite size is approximately 38–39 nm from 300 to 500 °C, increasing to 81 nm at 700 °C as its texture contribution decreases. These observations are consistent with an increased nucleation density or oxygen-limited coherent growth under high oxygen flux, in keeping with the reported sensitivity of β-Ta stability to oxygen [33].

### D. *Implications for the Growth of Ta on c-Plane Sapphire*

A central implication of this study is that an $O_2$ partial pressure as low as $2.5 \times 10^{-9}$ Torr can favor α-Ta (111) over α-Ta (110) on c-plane sapphire between 300 and 700 °C. Variations in residual oxygen-bearing gases between deposition systems may therefore contribute to the differing α-Ta orientations reported under nominally similar growth conditions [6,10,11,29,78]. Reporting the residual $O_2$ and $H_2O$ partial pressures, where measurable, would improve comparison of growth procedures. The influence of $H_2O$ was not isolated in this study, and it remains unknown whether water vapor or other oxygen-bearing species produce the same orientation response as molecular oxygen. The growth map identifies three distinct process regimes without seed layers: α-Ta (110)-dominant growth from the UHV baseline over $T_h = 0.17$–$0.36$; strongly α-Ta (111)-dominant growth at $2\Phi_{O_2}/\Phi_{Ta} \approx 2 \times 10^{-4}$ over $T_h = 0.17$–$0.30$; and β-Ta-dominant growth at $2\Phi_{O_2}/\Phi_{Ta} = 19$ and $T_h = 0.17$. Because microwave loss in Ta-on-sapphire devices has been associated with the epitaxial metal–substrate interface [4], the ability to modify the preferred α-Ta orientation through a controlled trace oxygen supply provides a route to test the relationship between interfacial crystallography and superconducting device loss.

## V. CONCLUSIONS

In this study, we mapped oxygen-dependent phase and orientation selection in 100 nm Ta films deposited on c-plane sapphire by DC magnetron sputtering. Substrate temperatures ranged from 300 to 900 °C, while oxygen supply spanned 6.6 orders of magnitude ($2\Phi_{O_2}/\Phi_{Ta} \leq 4.8\times10^{-6}$ to 19). Absorption-corrected texture coefficients referenced to a calculated α-Ta/β-Ta powder pattern enabled comparison of the α-Ta(110), α-Ta(111), and β-Ta(001) components. Without deliberate oxygen addition or seed layers, α-Ta(110) dominated at all temperatures, and elongated grains exhibited three preferred in-plane directions consistent with the rotational symmetry of c-plane sapphire. By contrast, a trace oxygen flux ratio of $2.2\times10^{-4}$ favored α-Ta(111), with 98–99% of the normalized ternary texture weight from 300 to 700 °C and 65% at 900 °C. The weak temperature dependence of this preference from 300 to 700 °C is consistent with orientation selection during nucleation, while its reduction at 900 °C may reflect increased grain-boundary mobility.

Substantial β-Ta(001) formed only at the highest oxygen supply ($2\Phi_{O_2}/\Phi_{Ta} = 19$). It was the only phase detected at 300 °C, dominated at 500 °C, became a minor component at 700 °C, and was undetected at 900 °C. The apparent β-Ta(001)/α-Ta(110) selection energy of 1.07 ± 0.16 eV agrees within uncertainty with the reported 1.10 eV activation energy for oxygen diffusion in Ta. Together with the estimated oxygen diffusion lengths relative to film thickness, this agreement supports a kinetic interpretation in which oxygen accumulation at the growth front favors β-Ta formation. Although the microscopic mechanisms remain to be resolved, these results establish oxygen supply as a sensitive control parameter for Ta phase and orientation selection and identify growth conditions for reproducible Ta films on c-plane sapphire.

# ACKNOWLEDGMENTS

This material was based upon work supported by the National Science Foundation under Grant No. 2328747.

# AUTHOR DECLARATIONS

**Conflicts of interest**

The authors have no conflicts to disclose.

**Author Contributions**

**Marius O. Eji:** Writing-Original draft (lead); Data curation (lead); Formal Analysis (lead); Conceptualization (lead); Investigation (lead). **Tyler J. Kirkpatrick:** Data curation (lead); Formal Analysis (lead); Writing-Original draft (support). **Ty J. Cawen:** Investigation (supporting). **Joseph P. Corbett:** Data curation (lead); Supervision (lead); Validation (lead); Conceptualization (lead); Investigation (lead); Writing-review & editing (lead); funding acquisition (lead).

# DATA AVAILABILITY

The data that support the findings are available upon request.

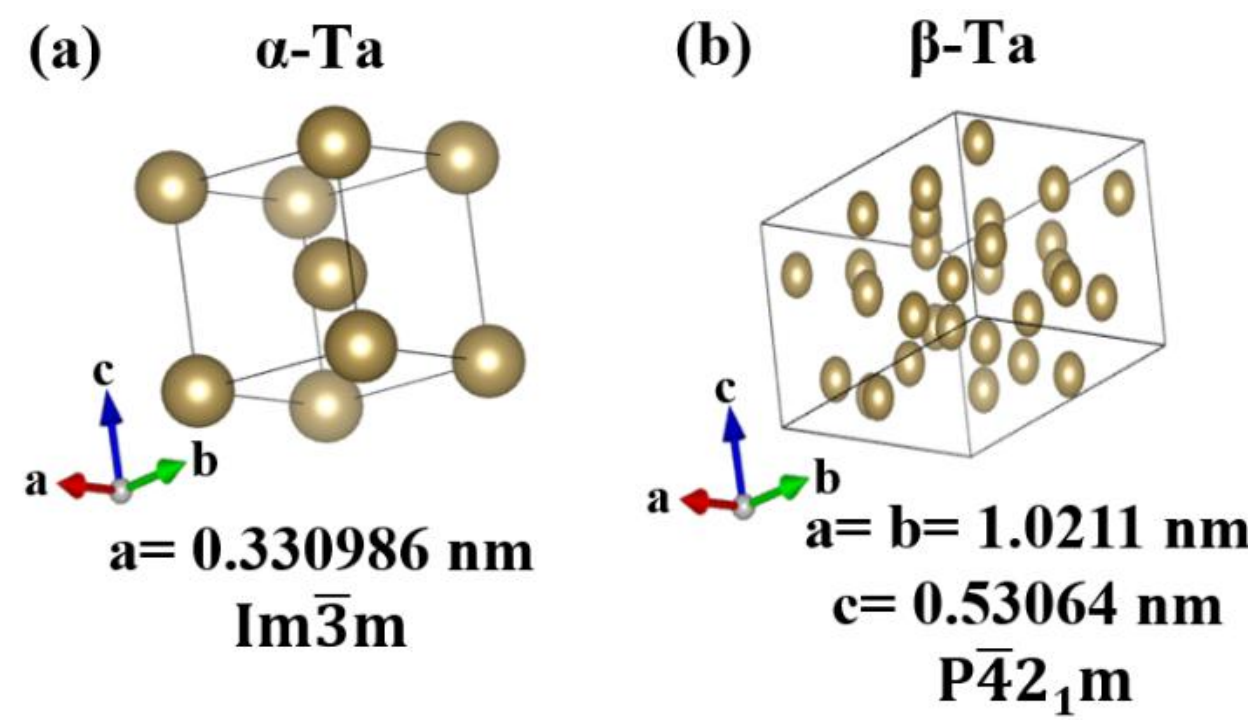

**Figure 1** Crystal models of the two phases of Ta. (a) α-Ta, body-centered cubic (Im$\bar{3}$m a = 0.330986 nm). (b) β-Ta, the tetragonal σ-type Frank–Kasper structure (P$\bar{4}2_1$m; a = 1.00211 nm, *c* = 0.5306 nm) with 30 atoms per unit cell

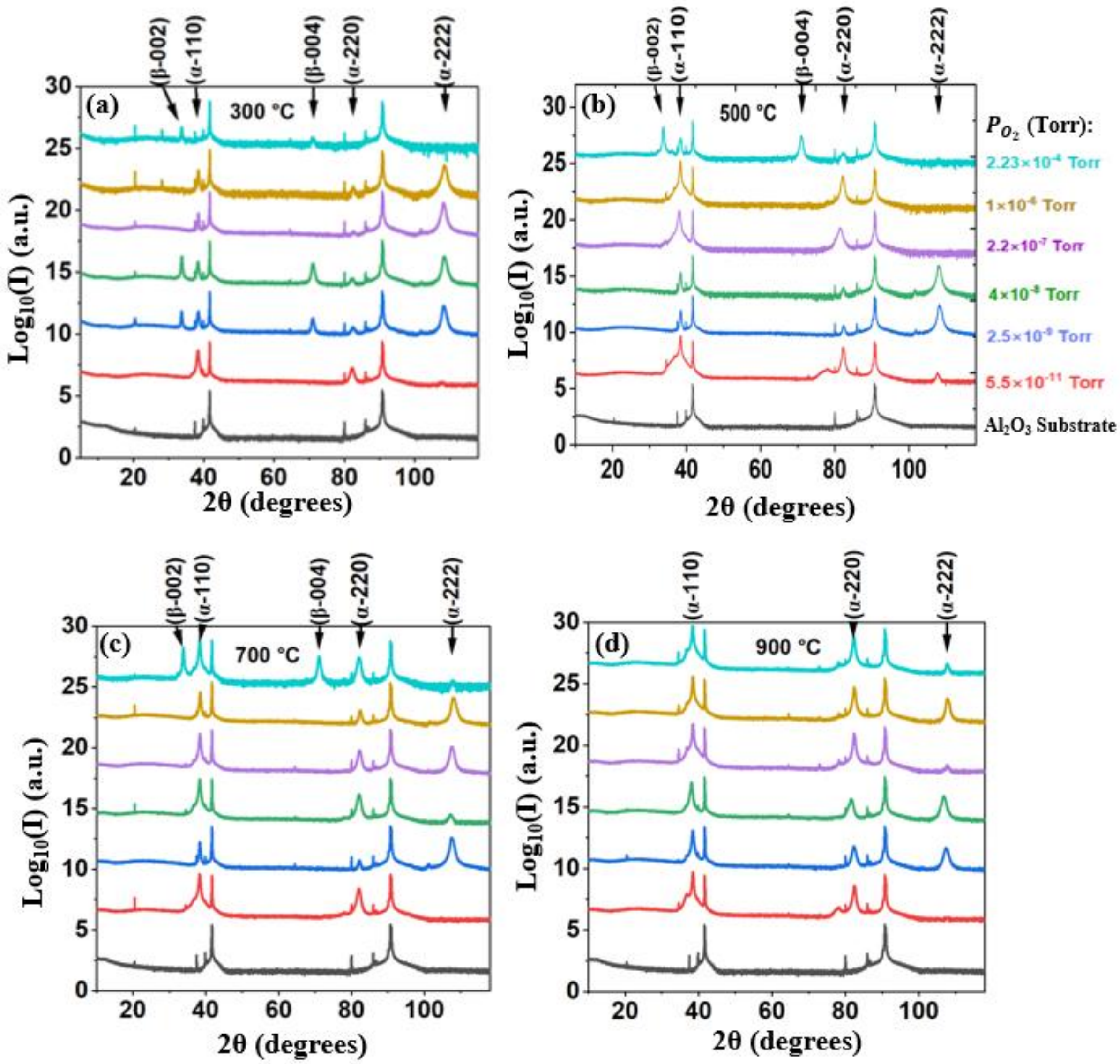

**Figure 2** Waterfall plots of XRD spectra spanning the growth conditions across deposition temperature and oxygen partial pressure. Oxygen conditions are color-coded where the image lines match the oxygen partial pressure. Each panel is a separate growth temperature. XRD diffractograms are plotted on a log scale and contain an arbitrary offset to enable a waterfall comparison. (a) growths at 300 °C, (b) growths at 500 °C, (c) growths at 700 °C, (d) growths at 900 °C

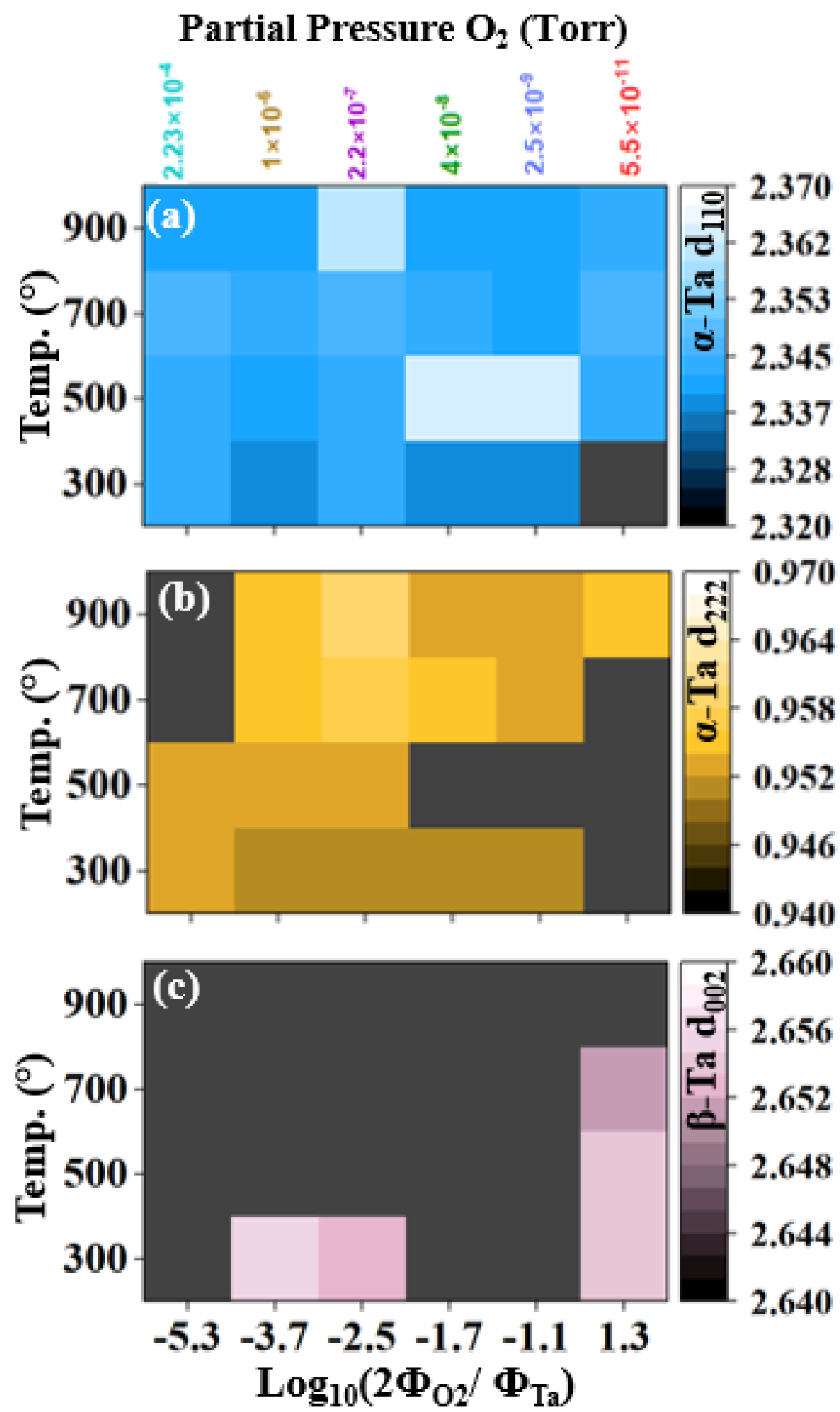

**Figure 3** Tiled heat maps of Ta d-spacing for both orientations and phases on c-plane sapphire. Each panel is a single phase/orientation from the leading XRD reflection (a) is α-Ta (110) (b) is α-Ta (222) (c) is β-Ta (002). The color intensity is the respective d-spacing value. Where the center color is set to the bulk reported values. Grey tiles are where no reflection was detected. The vertical axis indicates the temperature of the growth, the x-axis is the oxygen content.

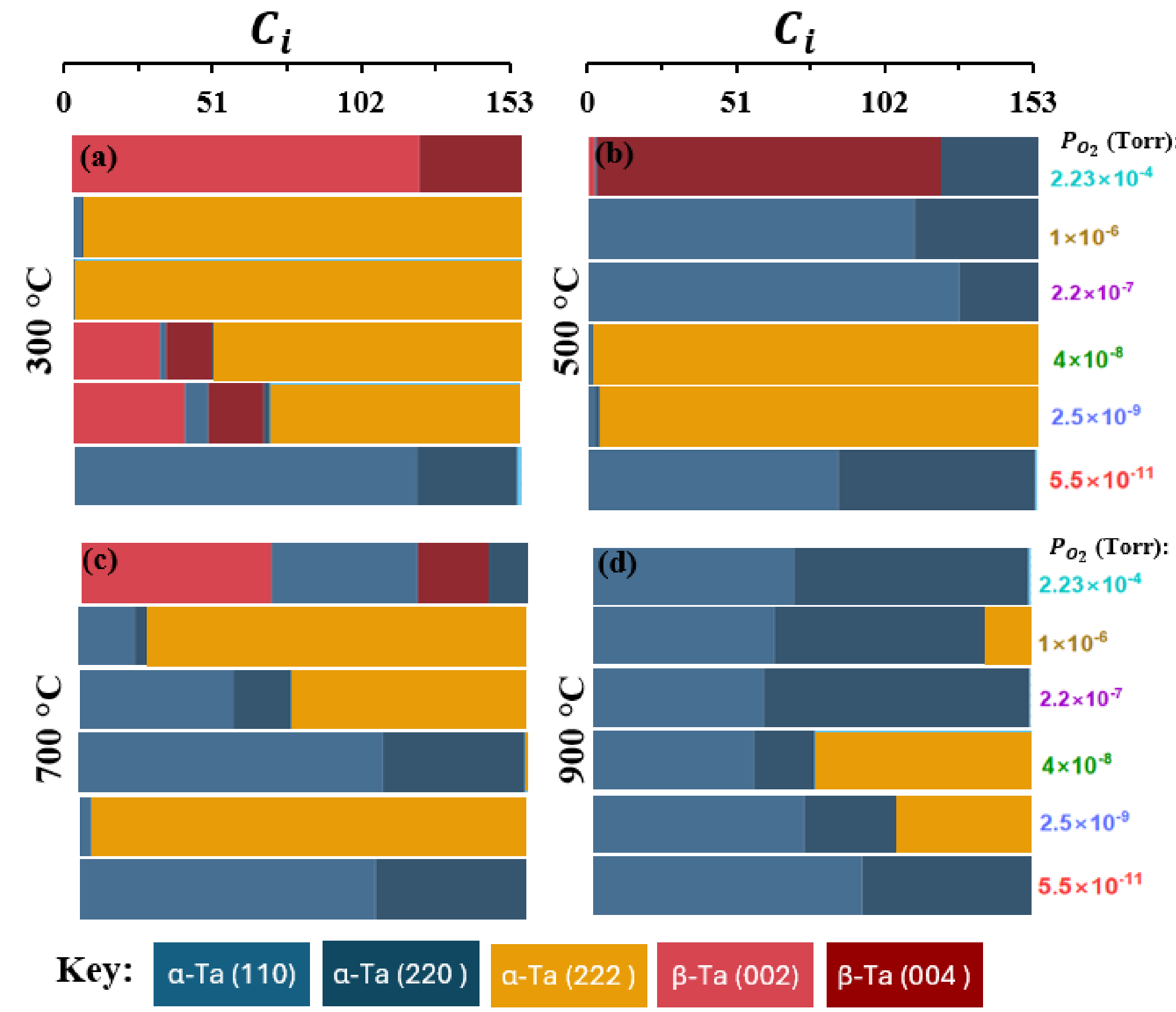

**Figure 4** Bar graphs of the texture coefficient spanning the spanning the growth conditions across deposition temperature and oxygen partial pressure. Oxygen conditions are color-coded where the image lines match the oxygen partial pressure. Each panel is a separate growth temperature. (a) growths at 300 °C, (b) growths at 500 °C, (c) growths at 700 °C, (d) growths at 900 °C

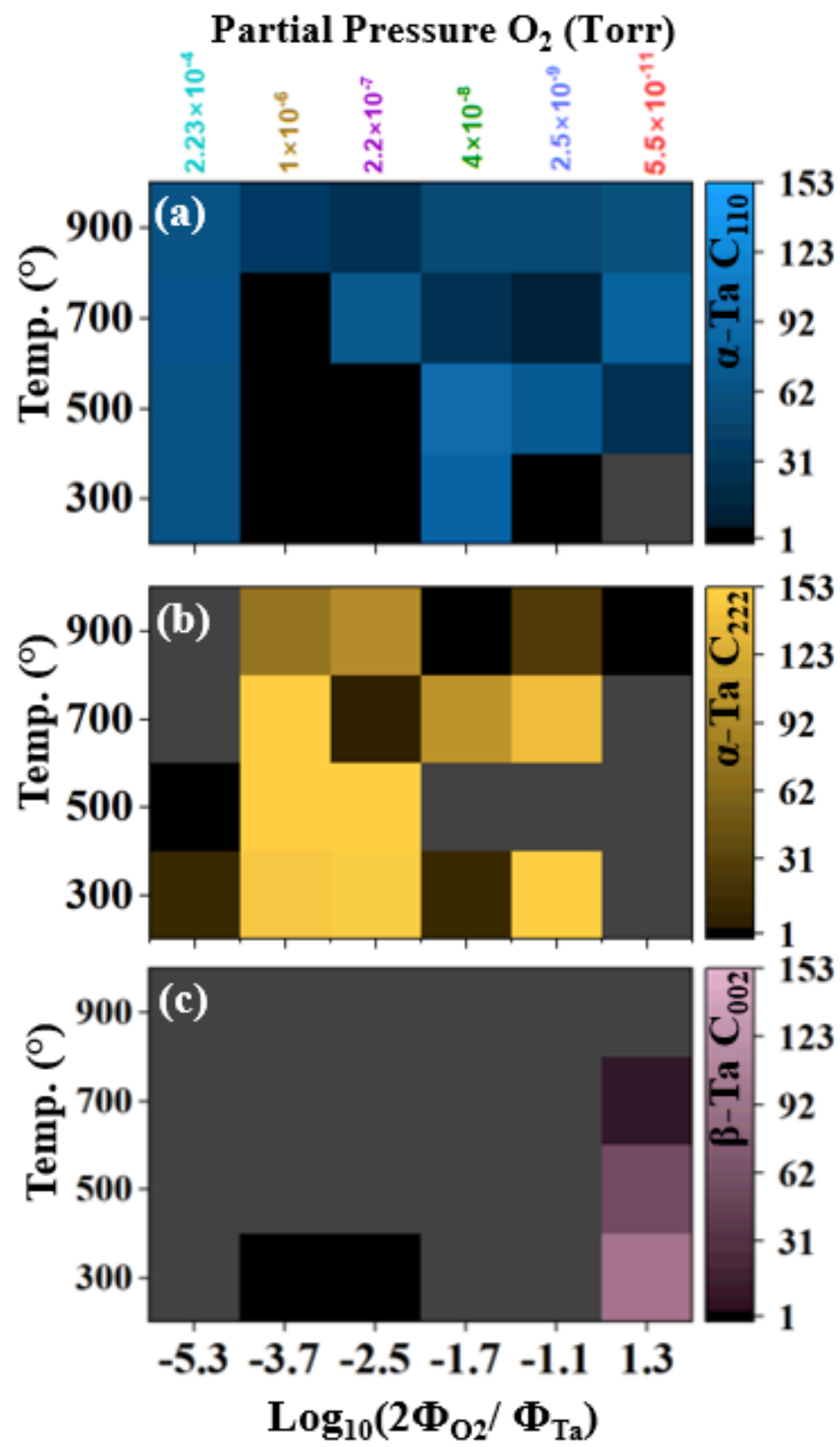

**Figure 5** Tiled heat maps of Ta orientations and phases on c-plane sapphire. Each panel is a single phase/orientation from the leading XRD reflection (a) is α-Ta (110) (b) is α-Ta (222) (c) is β-Ta (002). The color intensity is the respective texture coefficient. Black tiles are weak reflection C $\lesssim$ 1. Grey tiles are no reflection detected. The vertical axis indicates the temperature of the growth, the x-axis is the oxygen content.

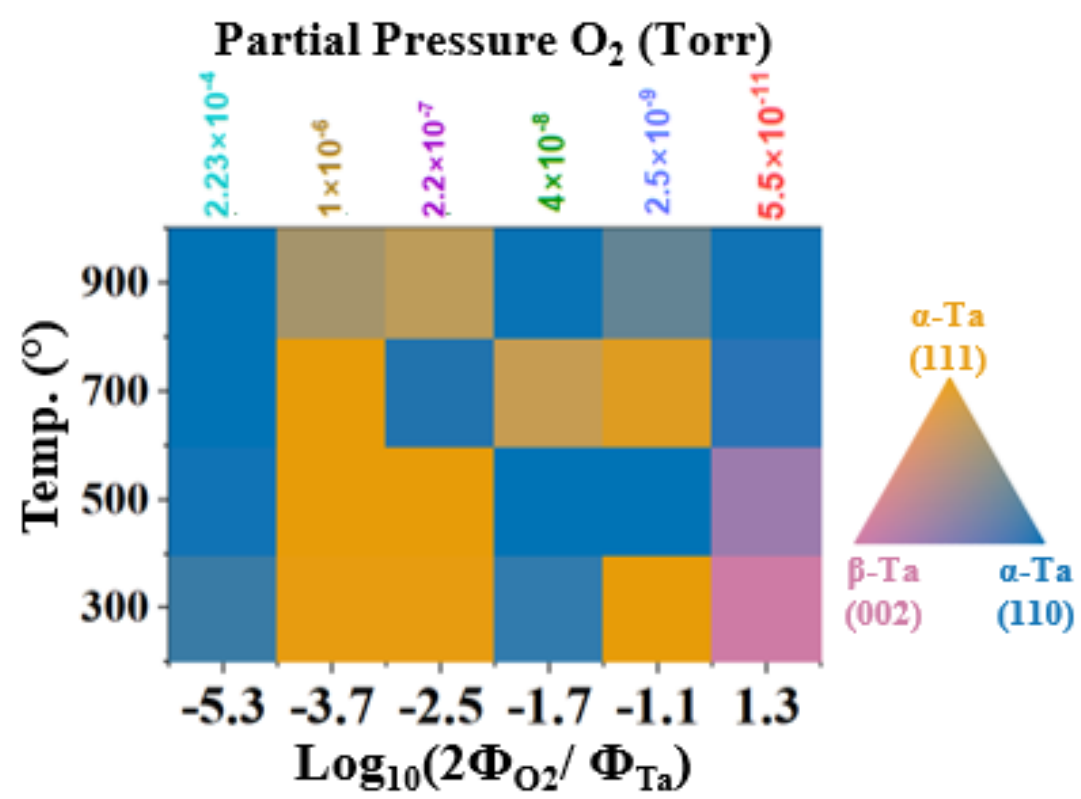

**Figure 6** Growth phase map of Ta on c-plane sapphire. Each tile is one film, colored by its position on the ternary diagram (right) of the α(110), α(111), and β(001) family texture coefficients; open circles on the ternary mark the 24 films. Bottom axis: oxygen-to-tantalum flux ratio, $\log_{10}(2\Phi_{O_2}/\Phi_{Ta})$, at a Ta deposition rate of 1.49 nm $s^{-1}$, which transfers to other deposition systems (the leftmost column, no added $O_2$, is an upper bound set by the base pressure). Top axis: the pre-deposition $O_2$ partial pressure set in this chamber. Left axis: substrate temperature; right axis: $T_h = T/T_m$. Tile labels name components with fractional weight $\geq 0.10$.

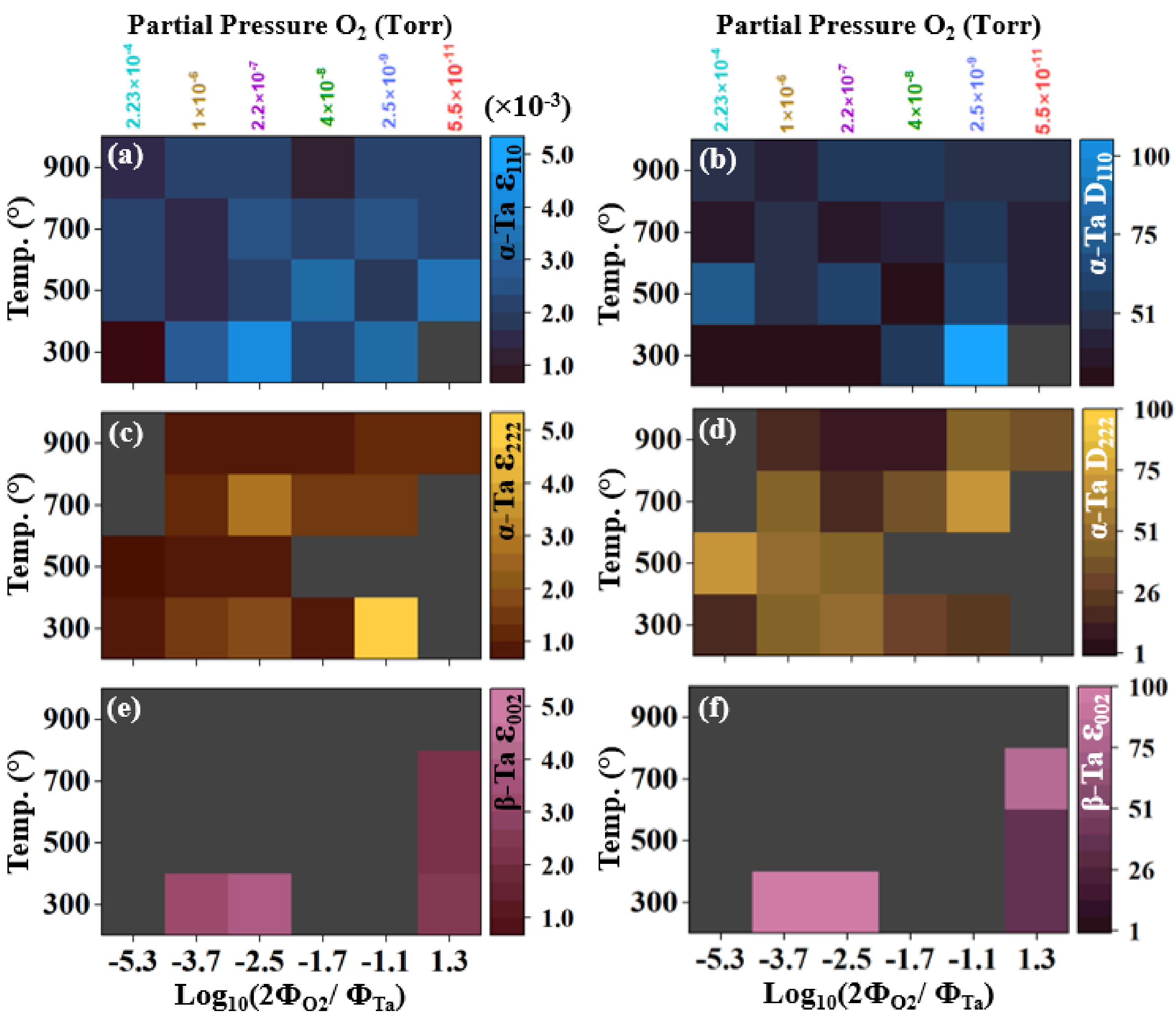

Figure 7 Out-of-plane crystallite size D and microstrain ε (in units of $10^{-3}$) of the (a,b) α-Ta (110), (c,d) α-Ta (111), and (e,f) β-Ta (001) components respectively, obtained from the 110, 222, and 002 reflections, on the axes of Fig. 4. Each component is shown on a sequential single-hue scale in the color of its ternary vertex; blank tiles mark conditions where the component is not detected. Uncertainties are ±20 nm and $\pm0.75\times10^{-3}$.

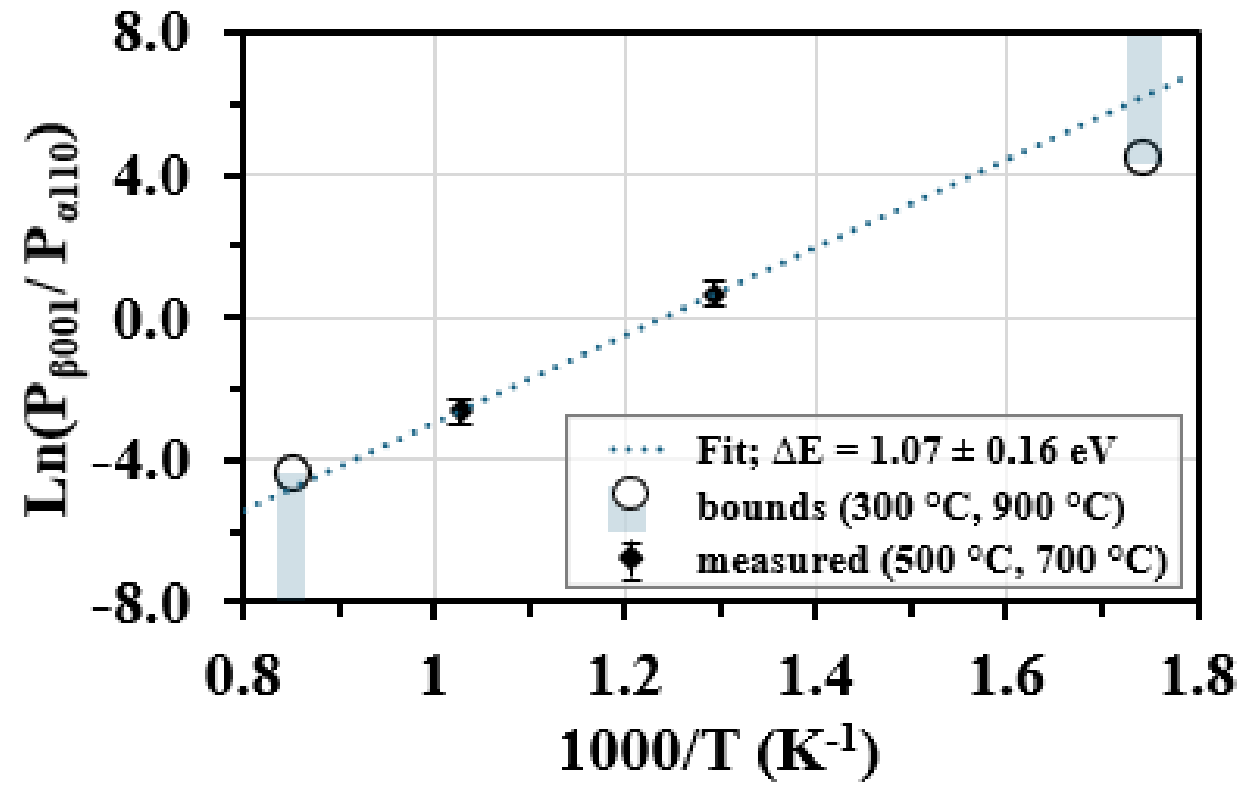

**Figure 8** Arrhenius plot of $\ln(P_{\beta 001}/P_{110})$ against 1000/T at $2\Phi_{O_2}/\Phi_{Ta}$ = 19. Filled circles: mixed-phase films at 500 and 700 °C (error bars ±0.35), with the fitted line, ΔE = 1.07 ± 0.16 eV. Open symbols with arrows: one-sided limits for the single-phase films at 300 °C (lower bound) and 900 °C (upper bound), in which one component is not detected; the undetected coefficient is replaced by that of a reflection with an integrated area of 1.2, on the same scale as the measured areas and equal to that of the weakest reflection detected in this study. Error bars: ±25% per coefficient. Top axis: growth temperature in °C.

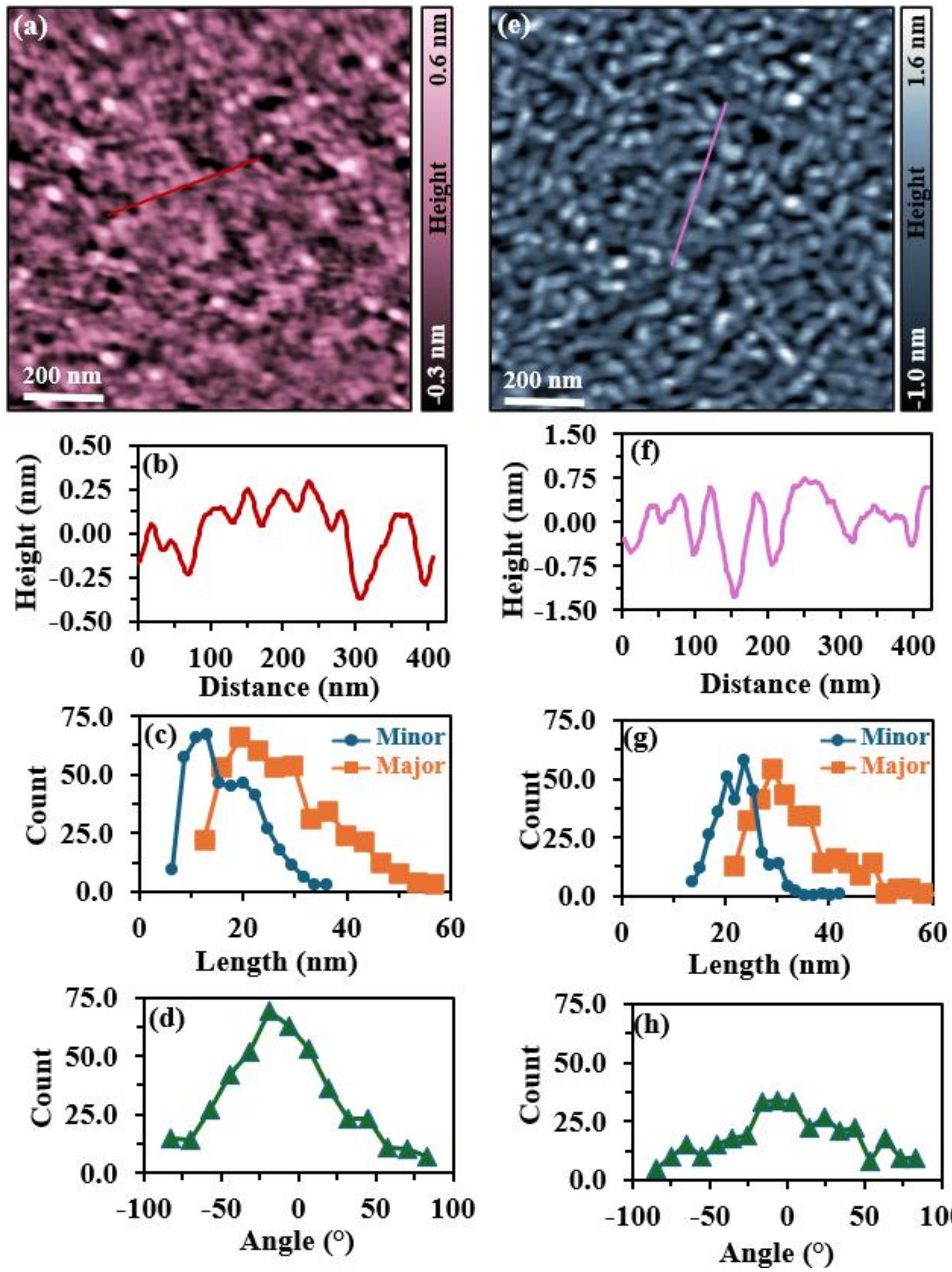

**Figure 9** AFM topographs of (a) a β-Ta film and (d) an α(110) film, with (b, e) the corresponding segmented grain masks and (c, f) the distributions of the major-axis angle of the fitted ellipses. Height scales: ±0.25 nm (a) and ±0.75 nm (d).